# ENHANCE ACCURACY IN SOFTWARE COST AND SCHEDULE ESTIMATION BY USING "UNCERTAINTY ANALYSIS AND ASSESSMENT" IN THE SYSTEM MODELING PROCESS

*Kardile Vilas Vasantrao, Department of Computer Science.*
*Tulajaram Chaturchand College, Baramati.*
*Affiliated by Pune University, Pune, Maharashtra ( India)*

## Abstract

*Accurate software cost and schedule estimation are essential for software project success. Often it referred to as the "black art" because of its complexity and uncertainty, software estimation is not as difficult or puzzling as people think. In fact, generating accurate estimates is straightforward—once you understand the intensity of uncertainty and framework for the modeling process. The mystery to successful software estimation—distilling academic information and real-world experience into a practical guide for working software professionals. Instead of arcane treatises and rigid modeling techniques, this will guide highlights a proven set of procedures, understandable formulas, and heuristics that individuals and development teams can apply to their projects to help achieve estimation proficiency with choose appropriate development approaches*

*In the early stage of software life cycle project manager are inefficient to estimate the effort, schedule, cost estimation and its development approach .This in turn, confuses the manager to bid effectively on software project and choose incorrect development approach. That will directly effect on productivity cycle and increase level of uncertainty. This becomes a strong cause of project failure. So to avoid such problem if we know level and sources of uncertainty in model design, It will directive the developer to design accurate software cost and schedule estimation, which are essential for software project success. However once the required efforts have estimated, little is done to recalibrate and reduce the uncertainty of the initial estimates.*

*This paper demonstrates terminology and typology of uncertainty is presented together with a framework for the modeling process, Brief reviews have been made of 14 different (partly complementary) methods commonly used in uncertainty assessment its interaction with the broader system development process and the role of uncertainty at different stages in the modeling processes.. The applicability of these methods has been mapped according to purpose of application, stage of the modeling process and source and type of uncertainty addressed.*

## I. Introduction

Design and chose approach is repeated incident in our daily life when we plan to go to our work .We estimate the time and risk need for design approach. The estimated time and risk fluctuates according external uncertain factor and theme's condition.

In our everyday life, we enhance our estimation based on past experience and historical data.

Design of software design approach is crucial because of today's dynamic environment of software development firm. The World Wide Web has provided a platform for companies to communicate and transact directly with their customers and partners. Challenges arise due to fast evolving technology and increased competition as companies are under constant pressure to develop new functionalities to satisfy changing client needs and to deliver them in short cycles at low costs. (Iansiti and MacCormack 1997).

Existing methods for software development have lot of options which can be classified into two categories, plan-driven (traditional) and practice–driven (Boehm and Turner 2003; Iansiti and MacCormack 1997). During the early stage of the plan-driven approach, the user needs are identified; requirements for new functionalities are specified; technical specifications are created; development processes are defined; specific project targets are spelled out; and, acceptance criteria and tests are outlined. Many of the CMM or ISO based methods belong to this category. The focus of the project team is on development and implementation according to the plan. As a result, the success of a project using plan-driven approach hinges on the validity and reliability of the project plan. On other hand if requirement of user in changeable form then it will increases level of uncertainty the plan driven approach is less effective.

In a dynamic environment such as the World Wide Web, it is difficult to predict technological changes, clients' needs and thus the solutions to clients' problems. Very often, the client lacks the knowledge and experience to describe their needs, not to mention requirements and specifications. There is evidence showing that effective learning by the client during the initial design phases of a project impacts positively on its success (Wastell 1999; Majchrzak et al 2005). Requirement changes at later stages, which often happen as a result of client learning occurring late in a project, could lead to not meeting client expectations, and budget or schedule overruns (Boehm 1989). Therefore, a rigid project plan becomes less effective in guiding the project team and could result in an obsolete system. Instead, the project team needs to collaborate with the client to help each other learn fast and the development process should be flexible enough to allow frequent changes.

The practice–driven (Agile) methodology refers to a family of software development methods that focus on customer needs, shorter development cycle time and adapting to changes requirements. The family includes Extreme Programming (Beck et al 2002), Scrum, DSDM and Crystal Family (Fowler 2005). Agile method is characterized by a chaotic perspective, collaborative values and principles, and barely sufficient technology (Highsmith 2002). Its core values include: individuals and interactions over processes and tools; working software over comprehensive documentation; customer collaboration over contract negotiation; responding to change over following a plan (Manifesto for Agile Software Development). Compared with the plan-driven approach, the agile methodology addresses the lack of knowledge of both the client (on the technology and development process) and the developer (on client's business needs) by encouraging closer collaboration. Usually, client representatives are collocated and work alongside the project team. Instead of working against the plan, frequent changes are embraced to address the client's changing business needs. As a result the success of project using practice –driven approach hinges on the effective communication skill and correct abstraction formalization. Of development Team. In such situation development practitioner and Organization and company has confuse for which approach is abandoning or which adopting because of the strength and weaknesses which will force to learner for accept *"Technology never fail it will fail to produce best result due to opponent opportune."*

So if developer get to know at early stage of estimation that which methodology is suited for which module of system then we are able to reduce failure cause of software process.

The aim of this paper to understand modeling process and uncertainty appearance in particular module for better to estimation process. In section II introduce various factors which affect software productivity. In section III Available software development approaches comparatives. In section IV Modeling process. In section V Methodologies for uncertainty assessment. In section VI Guide to select an appropriate methodology for uncertainty assessment In section VII Discussion and conclusions

## Section II

Problem: Design a Process**:** In software development, Modern "lightweight" methodologies are gaining ground on more traditional "heavyweight" methodologies. Both have their advantages and disadvantages, and appropriateness where we get best result. Many project fail because of inaccurate handling design approach or decision made early some time prove to be wrong later on. The most critical and crucial part of software development approach is when planning of design development is required in the early stage of the software life cycle where problem to be solved. Estimated, requirement by user is not completely understand and problem to be solved had not yet been completely revealed. The Major issue that separates the various processes that we looked at is the amount of up-front Planning they require. We can think of this as a spectrum, which at one end has a purely Plan oriented and other end practice oriented question is then for any given situation how do find the right approach.

Alternatively You can find a situation where the approach will give best result for this causes to handle this uncertainty we must be understand Risk handle strategies of each approach. Software Risk although there has been considerable debate about proper definition for ware risk, There is general agreement that risk always involves two characterizes:

Uncertainty and Complexity (Project Risk, Technical Risk, Business risk, .etc). Which will directly affect the Software deployment Process and approach? There are four broad control factors. This factor s is interconnected .when one changes at least one other factor must also change.

- Cost- or Effort. Available money impact the amount of effort put into the system
- Schedule – A Software project is impacted as the timeline is changed.
- Requirements-The scope of the work that needs to be done can be increased or decreased to affect the project.
- Quality – Cut control by reducing quality.

To avoid such problem if we know level and sources of uncertainty in model design, initial phase of development , It will directive the developer to design accurate software cost and schedule estimation. Which are essential for software project success . However once the required efforts have estimated, little is done to recalibrate and reduce the uncertainty of the initial estimates.

## Section –III

This comparison Focus on : Practice driven is sometimes characterized as being at the opposite end of the spectrum from "plan-driven" or "disciplined" methods. This distinction is misleading, as it implies that agile methods are "unplanned" or "undisciplined". A more accurate distinction is that methods exist on a continuum from "adaptive" to "predictive". Practice-driven lie on the "adaptive" side of this continuum.

Adaptive methods focus on adapting quickly to changing realities. When the needs of a project change, an adaptive team changes as well. An adaptive team will have difficulty describing exactly what will happen in the future. The further away a date is, the vaguer an adaptive method will be about what will happen on that date. An adaptive team can report exactly what tasks are being done next week, but only which features are planned for next month. When asked about a release six months from now, an adaptive team may only be able to report the mission statement for the release, or a statement of expected value vs. cost.

Predictive methods, in contrast, focus on planning the future in detail. A predictive team can report exactly what features and tasks are planned for the entire length of the development process. Predictive teams have difficulty changing direction. The plan is typically optimized for the original destination and changing direction can cause completed work to be thrown away and done over differently. Predictive teams will often institute a change control board to ensure that only the most valuable changes are considered.

**Table 1** :Categories wise best practice

| Plan Driven Software Development | Practice –driven Software Development |
|---|---|
| High Criticality | Low Criticality |
| Junior Developers | Senior Developer |
| Requirements do not change Often | Requirement change often |
| Large number of Development | Small number of developer |
| Culture that demands order | Culture that thrives on chaos |

**Section –IV: Modeling as a part of Project planning in system development** is one of the most critical activities within the project lifecycle. **Project plan development** is the main part of Project Planning Stage. The project manager takes the responsibility for creating a project plan that is a formal document showing the basis upon which to assess the performance of the project and measure its results. Let's review the steps of **project plan development** in detail.

Create the Work Breakdown Structure.

To create a project plan, you will need to determine the *Work Breakdown Structure* (the acronym "WBS") for your project. The WBS is a detailed list of all the phases, activities and jobs required for successful project completion. The WBS becomes the foundation for your project plan as you can use it to identify the resources required to deliver each activity or task listed. The WBS allows you to design simple to-do lists and task lists and then assign them to members of the project team.

When developing a project plan, you should remember that the WBS also depicts the dependencies between tasks. You will need to identify how each task is associated with other tasks and what (internal or external) dependencies can be set. **Project plan creation** requires setting clear milestones so be sure you have added milestones to your WBS sample.

1.    Define the Required Resources.

Once the tasks and activities required to deliver your project are set, your next step in creating a project plan is to define the resources required to do each task and activity. In your WBS showing the scope of the project you should add a section that describes which resources are required and in which quantities and measures. At this step of **project plan development** the project resource base will be defined and types of resources will be identified.

Your project may require such resources as the following:

- Full-time and part-time employees
- Equipment and materials

Your task is to calculate how many people you should employ to do your project and then define basic suppliers who will provide equipment and materials. In your WBS you need to specify this information.

2.    Design a Project Schedule.

After the Work Breakdown Structure is completely outlined, your next step in creating a project plan is to schedule tasks and define durations for activities listed in the WBS. You will need to create a *project schedule* that shows the task execution sequence and sets due dates per activity within your project.

To build a project schedule the following information is necessary:

- Identified tasks and activities and their dependencies (both internal and external)
- Assignments to members of the project team (who will do which task)
- Risk mitigation strategies and a contingency plan    Critical milestones
- Allocated resources required for the project

At two previous **project plan development** steps this information has been identified so you can design a project schedule.

A typical modeling study will involve the following four different types of actors:

*Organization environment:* the Organization Environment responsible for the management of the software project, and thus of the modeling study and the outcome (the problem owner).

*System Designer and Analyst* : a person or an organization that develops the model and works it, conducting the modeling study. If the modeler and the Project manager belong to different organizations, their roles will typically be denoted consultant and client, respectively.

*System Testing*. a person that is conducting some kind of external review of a modeling study. The review may be more or less comprehensive depending on the requirements of the particular case to match the modeling capability of the modeler.

*End User or stakeholder*: an interested party with a stake in the system development issue, Stakeholders include the following categories: (1) competent authority (typically the Project manager, cf. above); (2) interest groups; and (3) general user or client.

The modeling process may be different according to the organization.

**Figure 1.**

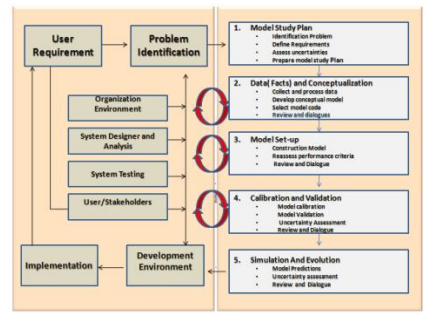

### Step 1

*Model study plan*. This step aims to agree on a Model Study Plan comprising answers to the questions: Why is modeling required for this particular model study? What is the overall modeling approach and which work should be carried out? Who will do the modeling work? Who should do the technical reviews? Which stakeholders/public should be involved and to what degree? What are the resources available for the project? The system designer and analyst needs to describe the problem and its context as well as the available data. A very important (but often over-looked) task is then to analyze and determine what are the various requirements of the modeling study in terms of the expected accuracy of modeling results. The acceptable level of accuracy will vary from case to case. It should, therefore, be defined through a dialogue between the modeler, project manager and stakeholders/client . In this respect an a priori analysis of the key sources of uncertainty is crucial in order to focus the study on the elements that produce most information of relevance to the problem at hand.

**Step 2**
_Data and conceptualization_ In this step the modeler should gather all the relevant knowledge about the study basin and develop an overview of the processes and their interactions in order to conceptualize how the system should be modeled in sufficient detail to meet the requirements specified in the model study plan. Consideration must be given to the spatial and temporal detail required of a model, to the system dynamics, to the boundary conditions and to how the model parameters can be determined from available data. The need to model certain processes in alternative ways or to differing levels of detail in order to enable assessments of model structure uncertainty should be evaluated. The availability of existing computer codes that can address the model requirements should also is evaluated.

**Step 3**
_Model set-up_. Model set-up implies transforming the conceptual model into a site-specific model that can be run in the selected model code. A major task in model set-up is the processing of data in order to prepare the input files necessary for executing the model. Usually, the model is run within a graphical user interface (GUI) where many tasks have been automated.

**Step 4**
_Calibration and Validation_. This step is con-cerned with the process of analyzing the model that was constructed during the previous step, first by calibrating the model, and then by validating its performance against independent field data. Finally, the reliability of model simulations for the intended type of application is assessed through uncertainty analyses. The results are described so that the scope of model use and its associated limitations are documented and made explicit.

**Step 5**
_Simulation and evaluation_. In this step the modeler uses the calibrated and validated model to make

**Section –IV**

4 Uncertainty terminology and classification(on the basis of Jens Christan Refesgard and his coauthor study of Uncertainty in the enviormental Modeling process-A framework and guidance)

**4.1. Definitions and taxonomy**
Uncertainty and associated terms such as error, risk and ignorance are defined and interpreted differently by different authors, see Walker et al. (2003) for a review. The different definitions reflect the underlying scientific philosophical way of thinking and therefore typically vary among different scientific disciplines. In addition they vary depending on theirpurpose. Some are rather generic, while others apply more specifically to model based software project management, such   By doing so we adopt a subjective interpretation of uncertainty in which the degree of confidence that a decision maker has about possible outcomes and/or probabilities of these outcomes is the central focus. Thus according to our definition a person is uncertain if s/he lacks confidence about the specific outcomes of an event. Reasons for this lack of confidence might include a judgment of the information as incomplete, blurred, inaccurate, unreliable, inconclusive, or potentially false. Similarly, a person is certain if s/he is confident about the outcome of an event. It is possible that a person feels certain but has misjudged the information (i.e. his/her judgment is wrong).

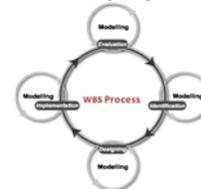

**Figure2:** The role of modeling in the software project management process

There are many different decision situations, with different possibilities for characterizing uncertainty. A first distinction is between ignorance as a lack of awareness that knowledge is wrong or imperfect, and uncertainty as a known degree of unreliability of knowledge, which translates into a state of confidence. In this respect Brown (2004) has defined taxonomy of imperfect knowledge as illustrated in Fig. 3. It is useful to distinguish between bounded uncertainty, where all possible outcomes are deemed 'known' and unbounded uncertainty, where some or all possible outcomes are deemed unknown. Since quantitative probabilities require 'all possible outcomes' of an uncertain event and each of their individual probabilities to be known, they can only be defined for 'bounded uncertainties'. If probabilities cannot be quantified in any undisputed way, we can still qualify the available body of evidence for the possibility of various outcomes in terms of plausibility or convincingness of the evidence (e.g. Weiss, 2003). If outcomes but no probabilities are known we have to rely on 'scenario analysis'.

The bounded uncertainty where all probabilities are assumed known (the lower left case in Fig. 3) is often denoted 'statistical uncertainty' (e.g. Walker et al., 2003). This is the case that is traditionally addressed in model-based uncertainty assessments. It is important to note that this case only constitutes one of many of the decision situations outlined in Fig. 3, and, in many situations, the main uncertainty in a decision situation cannot be characterized quantitatively.

**4.2. Sources of uncertainty**
Walker et al. (2003) describe uncertainty as manifesting itself at different locations in the model-based software project management pro-cess. These locations, or sources, may be characterized as follows: Context and framing, i.e. at the boundaries of the system to be modeled. The model context is typically determined at the initial stage of the study where the problem is identified and the focus of the model study selected as a confined part of the overall problem. This includes, for example, the external economic, environmental, political, social and technological circumstances that form the context of the problem. Input uncertainty in terms of external driving forces (within or outside the control of the software project manager) and system data that drive the model such as land use maps, pollution sources and climate data. Model structure uncertainty is the conceptual uncertainty due to incomplete understanding and simplified descriptions of modeled processes as compared to reality. Parameter uncertainty, i.e. the uncertainties related to parameter values. Model technical uncertainty is the uncertainty arising from computer implementation of the model, e.g. due to numerical approximations, resolution in space and time, and bugs in the software.

The total uncertainty on the model simulations, model output uncertainty, can be assessed by uncertainty propagation taken all the above sources into account.

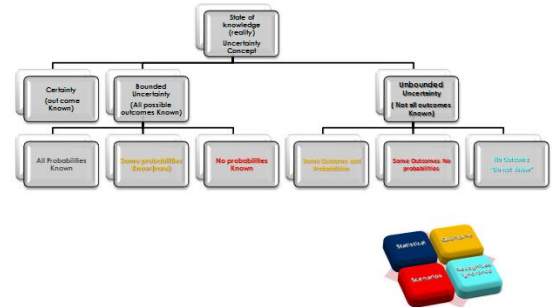

**Figure3** Taxonomy of imperfect knowledge resulting in different uncertainty situations (Brown, 2004).

**4.3. Nature of uncertainty**
Walker et al. (2003) explain that the nature of uncertainty can be categorized into:
Epistemic uncertainty, i.e. the uncertainty due to imperfect knowledge. Stochastic uncertainty or ontological uncertainty, i.e. uncertainty due to inherent variability, e.g. climate variability. Epistemic uncertainty is reducible by more studies, e.g. comprising research and data collection. Stochastic uncertainty is non-reducible. Often the uncertainty on a certain event includes both epistemic and stochastic uncertainty. An example is the uncertainty of the 100 year flood at a given site. This flood event can be estimated: e.g. by use of standard flood frequency analysis on the basis of existing flow data. The (epistemic) uncertainty may be reduced by improving the data analysis, by making additional monitoring (longer time series) or by deepening our understanding of how the modelled system works. However, no matter how perfect both the data collection and the mechanistic understanding of the system are, and, no mat-ter for how long historical data time series exist, there will al-ways be some (stochastic) uncertainty inherent to the natural system, related to the stochastic and chaotic nature of several natural phenomena, such as weather. Perfect knowledge on these phenomena cannot give us a deterministic prediction, but would have the form of a perfect characterization of the natural variability.

4.4. The uncertainty matrix
The uncertainty matrix in Table 1 can be used as a tool to get an overview of the various sources of uncertainty in a modeling study. The matrix is modified after Walker et al. (2003) in such a way that it matches Fig. 3 and so that the taxonomy now gives 'uncertainty type' in descriptions that indicate in what terms uncertainty can best be de-scribed. The vertical axis identifies the location or source of uncertainty while the horizontal axis covers the level and nature of uncertainty.

**Table2**: The uncertainty matrix (modified after Walker et al., 2003)

| Source of | Taxonomy | | | | Nat | |
|---|---|---|---|---|---|---|
| | Stat | Sce | Qu | Re | Epi | Stoc |
| Co Inp uts | Nat | | | | | |
| | Syst | | | | | |
| | Driv | | | | | |
| Mo del | Mod | | | | | |
| | Tec | | | | | |
| | Para | | | | | |
| Model | | | | | | |

It is noticed that the matrix is in reality three-dimensional (source, type, nature). Thus, the categories type and nature are not mutually exclusive, and it may be argued that the matrix should be modified in such a way that the two uncertainties within nature (epistemic and variability) should become sub cells within the type categories. This is not done for graphical reasons.

## Section –V

**5. Methodologies for uncertainty assessment**(on the basis of Jens Christan Refesgard and his coauthor study of Uncertainty in the enviormental Modeling process-A framework and guidance)

Many methodologies and tools suitable for supporting uncertainty assessment have been developed and reported in the scientific literature. We have selected 14 methods to repre-sent the commonly applied types of methods and tools. Guidance to the applicability of these methods is provided in Section 5. In the following the 14 methods are briefly reviewed in alphabetical order:

- ➥ Data uncertainty engine (DUE)
- ➥ Error propagation equations
- ➥ Expert elicitation
- ➥ Extended peer review (review by stakeholders)
- ➥ Inverse modeling (parameter estimation)
- ➥ Inverse modeling (predictive uncertainty)
- ➥ Monte Carlo analysis
- ➥ Multiple model simulation
- ➥ NUSAP
- ➥ Quality assurance
- ➥ Scenario analysis
- ➥ Sensitivity analysis
- ➥ Stakeholder involvement
- ➥ Uncertainty matrix

References to more detailed descriptions and to supporting software tools are provided in Refsgaard et al. (2005b). For several of the methodologies more extensive descriptions are available in the RIVM/MNP Tool Catalogue, that served as a starting point for the overview presented here (Van der Sluijs et al., 2004). A summary of statistically based methods for propagation of statistical uncertainty is given by Helton and Davis (2003).

## Section –VI

**6. Guide to select an appropriate methodology for uncertainty assessment**(on the basis of Jens Christan Refesgard and his coauthor study of Uncertainty in the enviormental Modeling process-A framework and guidance)

Some of the more important types of methodologies and associated tools that may be applied for assessing uncertainties were briefly reviewed above. The next question is which methodology should be selected for different purposes and in different situations. This is addressed from three different perspectives in the following three subsections.

6.1. Methodologies according to modeling process and level of ambition

Table 3 provides a list of applicable methodologies that are considered to be adequate at different stages in the modeling process. Furthermore, it includes hints for which methodologies are more suitable for comprehensive analysis with relatively large economic resources for the study and which methodologies correspond to a lower level of ambition (de-noted as ''basic'' in Table 3).

Uncertainty aspects are important throughout the modeling process. Considering the HarmoniQuA modelling protocol with the five steps shown in Fig. 1 and described in Section 2 above, uncertainty should be considered explicitly in all five modeling steps. However, it is treated in different ways at different stages of the modeling process. The three main actions of dealing with uncertainty may be characterized as:

**Identify and characterize sources of uncertainty**. The various sources of uncertainty need to be identified and characterized in Step 1 (model study plan). This should be done by the Project manager but typically after a dialogue with relevant stakeholders. Depending on the framing of the model study some of these uncertainties may be located as external non-controllable sources. It is crucial that uncertainty is considered explicitly so early in the definition phase of the model study. Here uncertainties are seldom quantified. It is also at this early stage that the first analyses are made on the acceptable level of uncertainty and the expected model performance.

**Reviews dialogue e decisions**. The last task in each of the modeling steps is a dialogue or decision task where a dialogue between project manager and modeler takes place. Often independent reviews are conducted as a basis for the decision and stakeholders and/or the general public are involved in the dialogue. As part of this dialogue, uncertainty aspects become important, e.g. when discussing whether there are sufficient data to proceed with the modeling, or whether the uncertainty of the model simulations is at a level where the results can be expected to be useful. The reviews and the stakeholder dialogues are also important platforms for a reflection on whether the assumptions made in the model are realistic and on how the study out-come may be influenced by the implicit and explicit assumptions made in the model. In many cases, more than one assumption is scientifically tenable. If such assumptions influence the model outcome, then the ignorance regarding which assumption is the best assumption can be an important source of uncertainty.

**Uncertainty assessment and propagation**. Towards the end of Step 4 an uncertainty analysis should be made of the calibration and validation results. This is used for evaluating

**Table3:** Suitable methodologies to deal with uncertainty at various stages of a modeling process

| Type of uncertainty aspect | Step in the modeling process (cf. Fig. 1) | Level of ambition/available resources | |
|---|---|---|---|
| | | Basic | Compreh. |
| Identify and characterize | Model study plan (Step 1) | UM | EPE, SI, UM |
| Reviews-dialogue-decisions | Review of Step 2 Review of Step 3 Review of Step 4 Review of Step 5 | QA | EPR, QA (Update of) UM |
| Uncertainty assessment and Propagation | Uncertainty analysis of calibration and validation (Step 4) | DUE, EPE, SA | DUE, EPE, EE, IN-PA, IN-UN, MCA, MMS, NUSAP, SA |
| | Uncertainty analysis of simulation (Step 5) | DUE, EPE, SA | DUE, EPE, EE, IN-UN, MCA, MMS, NUSAP, SC,SA, SI |

Abbreviations of methodologies: DUE, data uncertainty; EPE, error propagation equations; EE, expert elicitation; EPR, extended peer review (review by stake-holders); IN-PA, inverse modeling (parameter estimation); IN-UN, inverse modeling (predictive uncertainty); MCA, Monte Carlo analysis; MMS, multiple model simulation; NUSAP, NUSAP; QA, quality assurance; SC, scenario analysis; SA, sensitivity analysis; SI, stakeholder involvement; UM, uncertainty matrix.

**6.2. Methodologies according to source and type of uncertainty**
Table 5 provides a list of applicable methodologies for ad-dressing uncertainty of different types and originating from different sources. Note that the nature of uncertainty (epistemic or stochastic) has been omitted as compared to the uncertainty matrix in Table 1. The reason for this is that this is a third di-mension and that each of the cells below may be divided into reducible (epistemic) and irreducible (stochastic) uncertainty.

It is noted that none of the methods covers all the cells of the table, implying that for all modeling studies a suite of uncertainty methodologies has to be selected and applied. Some more general methods, such as expert elicitation, are potentially applicable for different types and sources of uncertainty, while other more specialized methods, such as Monte Carlo analysis, are only applicable for one type (here statistical uncertainty) and a couple of sources of uncertainty.

6.3. Methodologies according to purpose of use

The methodologies can roughly be divided in five groups that differ in purpose of use: possible biases in model imulations and assessing whether the model performance is good enough compared to the agreed accuracy requirements. Similarly, uncertainty analysis of simulations should be carried out in Step 5. Here the uncertainties in the problem framing (the context) and the management scenarios are also taken into account.

Methods for preliminary identification and characterization of sources of uncertainty. This category is identical to the first category in the first row in Table 2. The uncertainty matrix used together with stakeholder involvement is a suitable tool for this purpose. If a first rough quantification is desired the simple error propagation equations may be suitable.

Methods to assess the levels of uncertainty for the various sources of uncertainty. This use is addressed in some de-tails in in

Table 3. As can be seen many different methodologies may be suitable here. The exact selection will vary from case to case. It is noted from Table 4, that different methods apply to the different types of uncertainty (e.g. statistical versus qualitative uncertainty). Methods to propagate uncertainty through models. When all sources of uncertainty have been assessed they can be propagated through a model to assess the total uncertainty. In practice uncertainty propagation is often confined to include the data/parameters/model characteristics that have a significant effect on the total uncertainty. This selection is often supported by a sensitivity analysis. The methods suitable for uncertainty propagation are listed in the last row in Table 4. It is noted that uncertainty propagations is much easier to do for statistical and scenario uncertainty, while NUSAP and the simple error propagation equations are the only methods suitable for qualitative uncertainty (and ignorance). In practice uncertainty propagation of mixed statistical/qualitative uncertainty is very difficult to do in a rigorous manner. Methods to trace and rank sources of uncertainty. When the total uncertainty has been estimated it is often interesting to know how much the various sources contributed to the total uncertainty. This can be analyzed by used of Monte Carlo techniques and sensitivity analysis as far as the statistical uncertainty is concerned, while NUSAP may support such analysis with respect to the more qualitative aspects. Methods to reduce uncertainty. When an uncertainty assessment has been made it is often desired to evaluate if some of the uncertainty can be reduced. The part of the uncertainty that is epistemic may be reduced in different ways. The classical approach in natural science is to collect more data and carry out additional studies to gain more knowledge. For modeling studies quality assurance and extended peer reviews (stakeholder involvement in the modeling process) may reduce the uncertainties as well.

**Table 4:** Correspondence of the methodologies with the source and types of uncertainty distinguished in the uncertainty taxonomy (inspired by Van der Sluijs et al., 2004)

| Source of uncertainty | | Taxonomy (types of uncertainty) | | | |
|---|---|---|---|---|---|
| | | Statistical uncertainty | Scenario uncertainty | Qualitative uncertainty | Recognized ignorance |
| Context and Framing | Natural, technological, economic, social, political | EE | EE, SC, SI | EE, EPR, NUSAP, SI, UM | EE, EPR, NUSAP, SI, UM |
| Inputs | System data | DUE, EPE, EE, QA | DUE, EE, SC, QA | DUE, EE | DUE, EE |
| | Driving forces | DUE, EPE, EE, QA | DUE, EE, SC, QA | DUE, EE, EPR | DUE, EE, EPR |
| Model | Model structure | EE, MMS, QA | EE, MMS, SC, QA | EE, NUSAP, QA | EA, NUSAP, QA |
| | Technical | | | | QA |

|  | Param eters | IN-PA, QA | IN-PA, QA | QA | QA |
|---|---|---|---|---|---|
| Model output uncertainty (via propagation) |  | EPE, EE, IN-UN, MCA, MMS, SA | EE, IN-UN, MMS, SA | EE, NUSAP | EE, NUSAP |

The bottom row lists methodologies suitable for uncertainty propagation. Abbreviations of methodologies: DUE, data uncertainty engine; EPE, error propagation equations; EE, expert elicitation; EPR, extended peer review (review by stakeholders); IN-PA, inverse modeling (parameter estimation); IN-UN, inverse modeling (predictive uncertainty); MCA, Monte Carlo analysis; MMS, multiple model simulation; NUSAP, NUSAP; QA, quality assurance; SC, scenario analysis; SA, sensitivity analysis; SI, stakeholder involvement; UM, uncertainty matrix.

# Section –VII

**7. Discussion and conclusions**(on the basis of Jens Christan Refesgard and his coauthor study of Uncertainty in the enviormental Modeling process-A framework and guidance)

A terminology and typology of uncertainty is presented with the aim to assist the management of uncertainty in modeling studies for integrated System resources management. Because we focus on the use of model studies in decision making, we have adopted a subjective interpretation of uncertainty in which the degree of confidence that a decision maker has about possible outcomes and/or probabilities of these outcomes is the central focus. Other authors define the term uncertainty not as a property (state of mind) of the decision maker but as a property (state of perfection) of the total body of knowledge or information that is available at the moment of judgment. Uncertainty is then seen as an expression of the various forms of imperfection of the available information and depends on the state-of-the-art of scientific knowledge on the problem at the moment that the decision needs to be made (assuming that the decision maker has access to the state-of-the-art knowledge). The state of perfection view goes well together with a traditional plan basis, while our definition allows taking broader aspects of uncertainty, including those usually dealt with in Practice oriented disciple, into account. The broader view is necessary if we want to consider all aspects of modeling uncertainty when modeling is used as an element in the broader Project management process.

We have reviewed 14 methods for assessing and characterizing uncertainty. The 14 methods have been mapped against a framework for the modeling process, its inter-action with the broader Project management process and the role of uncertainty at different stages in the modeling processes.

Numerous methods that deal with uncertainty exist. The 14 methods we have included are by no means exhaustive, but in-tend to present a representative cross-section of commonly ap-plied methods covering the various aspects of uncertainty in System resources management. Many methods reported in literature naturally fall within one of the 14 'boxes', while others fall in between. An example of a method that does not fit well to our selection of methods is the generalized uncertainty likelihood estimation (GLUE) method (Beven and Binley, 1992; Beven, 2002). GLUE can be used both as a kind of calibration method or as an uncertainty propagation method. It is based on the concept of equi-finality and can be seen as a method having similarities in approach with three of the above 14 methods: Inverse modeling (parameter estimation), Monte Carlo analysis and multiple model simulation. Similarly many software tools have functionality corresponding to a couple of the 14 methods.

None of these methodologies is applicable to address all the different relevant aspects of uncertainty in the modeling in relation to System resources management. Most of the methods we have selected are complementary in approaches and con-tent. However, there are also some important overlaps. The best example of that is the quality assurance method that in reality is a framework within which some of the other methods, such as stakeholder involvement and extended peer review are typically recommended. In the quality assurance tool MoST (Refsgaard et al., 2005a; Scholten et al., 2007) all other methods are incorporated.

The key conclusion of the analysis in this paper is that uncertainty assessment is not just something to be added after the completion of the modeling work. Instead uncertainty should be seen as a red thread throughout the modeling study starting from the very beginning. Traditionally, uncertainty assessments are carried out only at the end of a modeling study when the models have been calibrated and validated.

Standard techniques, often included in the model GUIs, are then used to propagate and quantify the uncertainty, e.g. sensitivity analysis or Monte Carlo analysis. The major argument towards this type of uncertainty assessments is that the standard techniques do typically only address one type of uncertainty, namely the statistical uncertainty. By performing the uncertainty analysis as an 'add-on' by standard techniques in the end of the model study, and report this as the uncertainty analysis, it is implicitly assumed that the statistical uncertainty is the most important uncertainty. The statistical uncertainty does, how-ever, only comprise a limited space of the total uncertainty, as illustrated in Fig. 3. Moving towards the use of models in a broader perspective, such as Project management plans and the participatory processes in the WBS, other types of uncertainty emerged that have not traditionally been addressed in a model study. It is therefore crucial that the uncertainty assessment is introduced in the introductory phase and tracked throughout the model study and that the identification and characterization of all uncertainty sources are performed jointly by the modeler, the system manager and stakeholders in connection with the problem framing and identification of the objectives of the modeling study.

There is not much to conclude, This is early in our study, our hope is that a systematic look towards the impact of uncertainty at module level and pattern of software development approaches, which will useful to describe, express configuration and enact software engineering process for global software development, in way that respective specific allocated process which reduce causes of software failure.

It is therefore crucial that the uncertainty is introduced in the introductory phase and tracked throughout the model study and identification, characterization of all uncertainty sources are performed jointly by the modeler, The software project manager and stakeholders in connection with the problem framing and identification of the objectives of the modeling study, which will help to developer to choose the development approaches as per level of uncertainty.

## Acknowledgements

I like to express our sincere thanks to Dr. Murumkar C.V., Principal, Tuljaram Chaturchand College, Baramati. I would also like to thanks to all staff member of computer sci. Dept. who inspire us for this work.